\documentclass[a4paper,fleqn]{cas-dc}

\usepackage[numbers]{natbib}
\usepackage{gensymb}

\def\tsc#1{\csdef{#1}{\textsc{\lowercase{#1}}\xspace}}
\tsc{WGM}
\tsc{QE}
\tsc{EP}
\tsc{PMS}
\tsc{BEC}
\tsc{DE}

\begin{document}
\let\WriteBookmarks\relax
\def\floatpagepagefraction{1}
\def\textpagefraction{.001}
\shorttitle{Electronic structure of glassy Li$_3$ClO}
\shortauthors{Young Won Choi et~al.}

\title [mode = title]{Structural characterization and electronic structure of Li$_{3}$ClO glasses for solid-state 
Li-ion batteries}                      



\author[1]{Young Won Choi}
\ead{ywchoi@kth.se}
\cortext[cor1]{Corresponding author}

\address[1]{Applied Materials Physics, Department of Materials Science and Engineering, Royal Institute of Technology, Stockholm SE-100 44, Sweden}

\author[2]{Moyses Araujo}[]
\ead{Moyses.Araujo@kau.se}

\author[1,3,4]{Levente Vitos}[]
\ead{leveute@kth.se}

\author[1]{Raquel Liz\'arraga}[orcid=0000-0002-6794-6744]
\ead{raqli@kth.se}

\address[2]{Department of Engineering and Physics, Karlstad University, Karlstad, Sweden}

\address[3]{Department of Physics and Astronomy, Division of Materials Theory, Uppsala University, Box 516, SE-751 20 Uppsala, Sweden}

\address[4]{Research Institute for Solid State Physics and Optics, Wigner Research Center for Physics, Budapest H-1525, Hungary}

\begin{abstract}
Energy storage technologies that can meet the unprecedented demands of a sustainable energy system based on intermittent energy sources require new battery materials. We investigate high ionic conductors, Li$_3$ClO glasses. In the present work we use a first principles method to model the amorphous structure of the glass. We characterize the structure by means of radial distribution functions, radial disctributions functions and coordination numbers. We compare to their crystalline counterparts. The electronic structure of the glass is compared to that of the crystalline material. The band gap of the glass appears to be slighly reduced compared to that of the crystal. We also investigate the chemical stability of the glass against Li metal electrode. 
The electrochemical stability of the glassy electrolite is evaluated against Li metal.
\end{abstract}




\maketitle

\section{Introduction}
The lithium-ion battery (LIB) is the fastest growing and most promising battery
for a broad range of applications, e.g. portable electronics, robotics and all-electric vehicles. 
Its most common design consists of a graphite anode, a Li metal oxide cathode
and an electrolyte made of Li salt and an organic solvent.
The electrolyte provides the conductive medium, in which the
Li ions travel back and forth from one
electrode to the other during discharge and charge.
Presently, three types of LIBs are envisioned as the next generation Li-ion battery, namely,
metal-air, multi-valent cation and all-solid state batteries~\cite{Braga2014}. The latter is
a very promising battery that has many advantages~\cite{Wang2015,Manthiram2017} and it is currently considered
the next step on major OEM's roadmaps\cite{EurBa}.
Firstly, the all-solid-state battery can be made thinner and safer than conventional LIBs,
because it does not need a separator and
it does not use flammable liquid materials for the electrolyte that are susceptible to
dangerous leaks. Secondly, the solid-state electrolyte can withstand voltages up to 10 V which can enable
high voltage cathodes, and consequently increase the energy density capacity of the battery by 20-50 \%.
Moreover, solid electrolytes usually show excellent storage stability and
very long life\cite{Takada2013}.

The challenge of the all-solid-state battery is
to increase ionic conductivity to be
comparable or even better than that of conventional liquid electrolytes
(typically around $10^{-3}$ S/cm at room temperature\cite{Wang2015}).
Solid electrolyte materials exhibit rather low ionic conductivity ranging from $10^{-8}$ to $10^{-5}$
S/cm (halides), $10^{-5}$ to $10^{-3}$ S/cm (oxides), $10^{-7}$ to $10^{-3}$ S/cm (sulfides) 
and $10^{-7}$ to $10^{-4}$ S/cm (hydrides)~\cite{Manthiram2017}.
Currently, ceramics, sulfides, solid polymers, glasses and composites of ceramics or glasses
are materials proposed for solid-state electrolyte technology.
Glasses have been known for a while to be fast ionic conducting materials, however
research efforts have been concentrated mostly
on synthesis and optimization of the thermal properties~\cite{Kondo1992}. 

Recently, Braga et al. proposed and patented
a new all-solid state battery architecture based on Li$^{+}$ or Na$^{+}$ glassy electrolytes
\cite{Braga2014,Braga2016,Braga2017,Braga2018}.
One the major breakthroughs of this new battery is the fact that, it can be
realized using Na which is abundant, inexpensive and capable of promoting
sustainable use of raw materials. Furthermore,
it may be made without cobalt in the cathode, which is
an advantage because cobalt presents several unresolved issues regarding health
\cite{linna2004,Holmstrom2018} and irresponsible mining practices~\cite{guardian}.
This battery promises to provide 2-3 times the energy storage
capacity of a comparable LIB and most importantly it appears not to form
dendrites, which are associated to short circuits in LIBs.

In the present study we investigate the amorphous structure of Li$_3$ClO glass and its electronic structure and we compare with its crystalline counterpart. The paper is organized as following. In section 2 we discuss the methodology and computational methods used in this investigation. Section 3 presents the results and discussions. Finally in section 4 we present the conclusions of our study. 

\section{Methodology}

%

Models of the amorphous structure have been generated by means of the stochastic quenching method (SQ)
\cite{Holmstrom2009,Holmstrom2010}. The method is based on Density Functional Theory\cite{Hohenberg,Martin} 
and it has been used extensively in the past to study amorphous systems 
e.g. amorphous alumina~\cite{Lizarraga2011}, amorphous graphite~\cite{Holmstrom2011}, amorphous YCrO$_3$~\cite{Araujo2014,Lizarraga2012}, amorphous GdRE$_2$\cite{Lizarraga2016}, etc. The size of the cell was carefully tested and models with 100, 250 and 500 atoms were generated. We did not observed significant differences between the 250-atoms and 500-atoms cell.
The equilibrium density of the glass was determined to be 1.63 g/cm$^{3}$ which is approximately 18\% lower than the crystalline counterpart.  

\begin{figure}
	\centering
		\includegraphics[scale=.95]{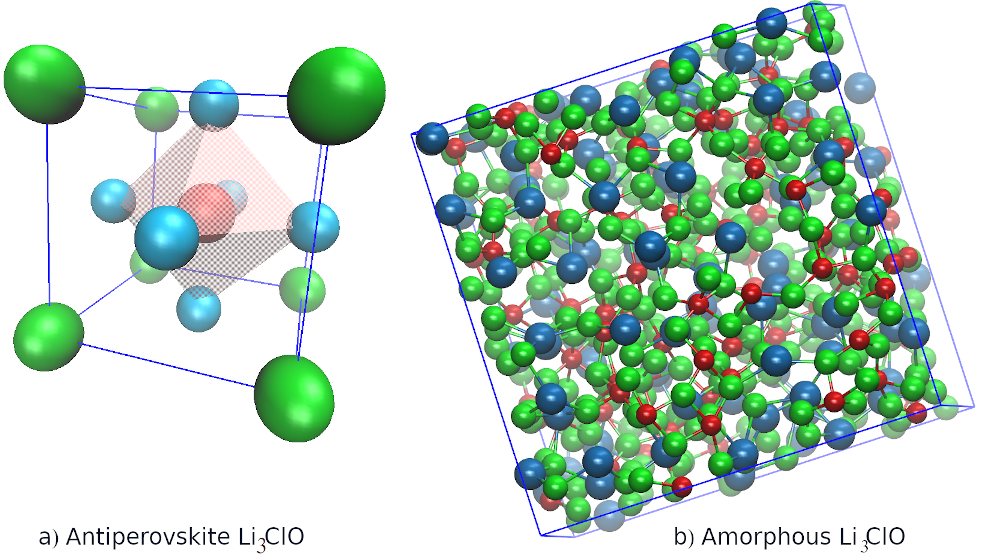}
	\caption{Antipervoskite Li$_3$ClO and Li$_3$ClO amorphous structures. The model of the amorphous structure
of Li$_3$ClO was obtained using SQ method and cell of 500 atoms. Red, blue and green balls represent Cl, O and Li atoms 
respectively.}
	\label{FIG:1}
\end{figure}

\subsection{Computational details}

The Vienna ab initio Simulation Package~\cite{vasp} (VASP)
was used during the quenching. The Perdew-Burke-Ernzerhof (PBE) implementation of the
exchage-correlation funational was used to generate the models of amorphous structures~\cite{Perdew1996} 
Due to the GGA's well known failure to describe the band gap correctly, we also performed 
GW0 and HSE calculations. The eigenstates of the electron wave functions
were expanded on a plane-waves basis set using pseudopotentials to describe the
electron-ion interactions within the projector augmented
waves approach (PAW)\cite{Blochl94}.

The convergence criterion for the electronic self-consistent cycle was fixed
at 10$^{-7}$ eV per cell and for the relaxation of the forces on
all ions, it was 10$^{-5}$ eV/\AA.
Calculations were performed at the $\Gamma$ k-point with a cutoff energy of 350 eV.
Calculations of antiperovskite structure were performed utilizing a k-point mesh of
$12\times12\times12$, except when density of states (DOS) was calculated, in which case a $22\times22\times22$ was used.  
Structural optimizations were performed by using a standard
conjugate gradient method during the stochastic quenching
procedure.

\section{Results and Discussion}

\subsection{Amorphous structure}

\begin{figure}
    \centering
    \includegraphics[scale=.27]{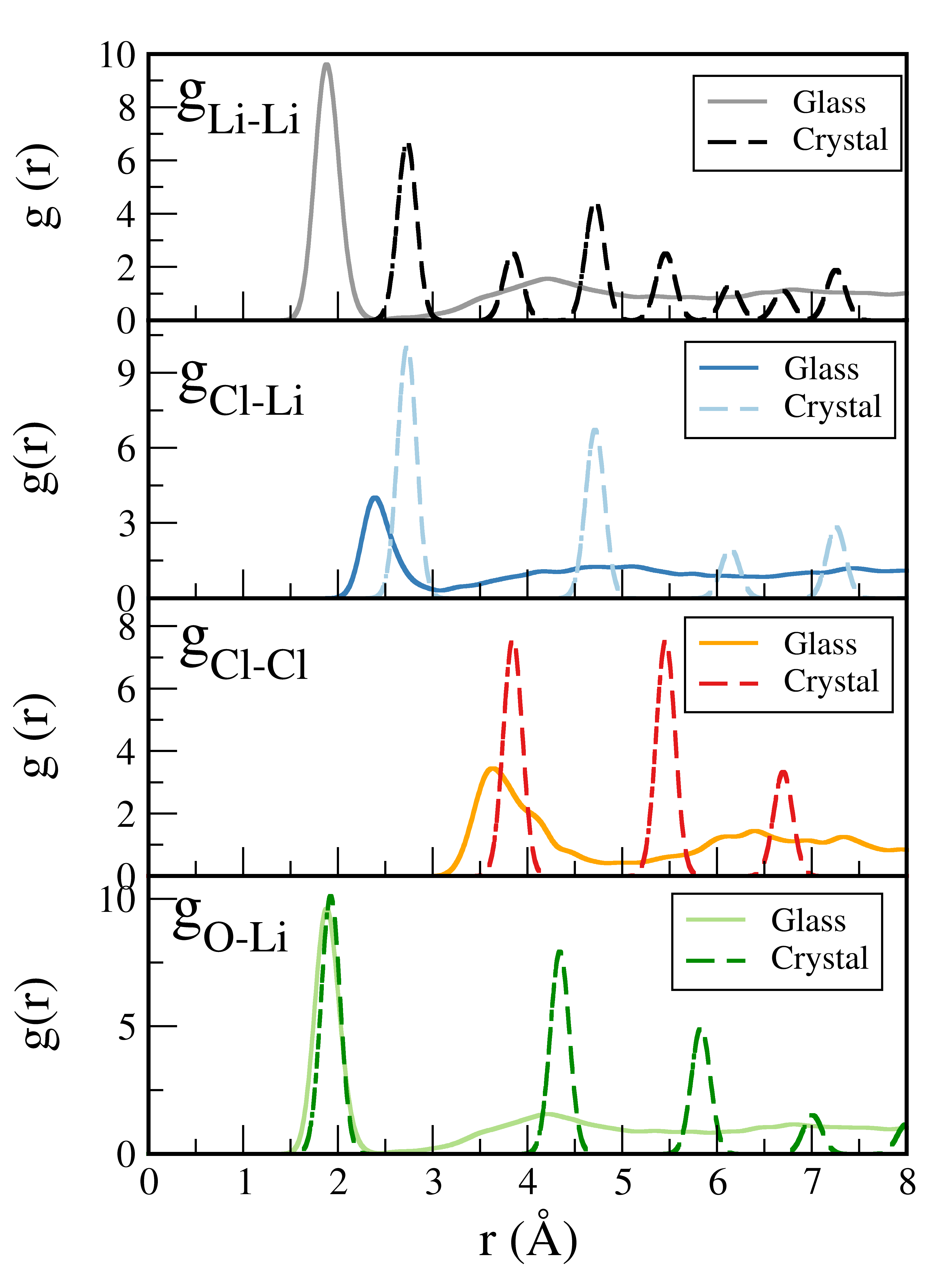}
    \caption{Radial distribution function of crystal and amorphous structure of Li$_3$ClO. (a) denote Li-Li, Cl-Li, Cl-Cl, and Cl-Li pairs and (b) O-O and Cl-O pairs.}
    \label{fig:RDFs}
\end{figure}

Fig.~\ref{fig:RDFs} shows selected radial distribution functions (RDFs)
for crystalline and amorphous Li$_3$ClO. In all cases, the first peak of the RDFs of the amorphous structure are shifted 
towards lower values than those of the crystalline structure. These is clearly reflected in the bond distances. 
Table~\ref{coord} lists the bond distances for both crystalline and amorphous structures. In all cases this distance is shorter
in the glass. In particular, the main peak of g$_{O-O}$ moves from 3.9 \AA in the crystalline structure 
to 3.1 \AA in the glss and a small peak develops at 1.5 \AA. This small peak has been associated 
to oxygen molecules impurities and 
have been observed before in other amorphous materials such as amorphous alumina~\cite{Arhammar2011}.

\begin{table}[width=.9\linewidth,cols=4,pos=h]
\caption{Bond distance in crystalline and amorphous Li$_3$ClO.}\label{bond}
\begin{tabular*}{\tblwidth}{@{} CCC@{} }
\toprule
Bond & \multicolumn{2}{c}{Shortest Distance (Å)}\\[.1cm]\cline{2-3}
& Crystalline & Amorphous\\
\midrule
O-O & 3.91 & 1.54\\
O-Li & 1.96 & 1.79\\
O-Cl & 3.39 & 3.25\\
Li-Li & 2.76 & 2.18\\
Li-Cl & 2.76 & 2.31\\
Cl-Cl & 3.91 & 3.41\\
\bottomrule
\end{tabular*}
\end{table}

Coordination numbers are listed in Table~\ref{bond}.
It is shown that most of pairs have smaller coordination number in amorphous state compared to crystalline one. 
For example, the coordination number of O-O pair is reduced from 6 into 4.4 by amorphization. 
However, O-Li coordination number increases from 6 to 10.25, which means O atom went to Li atom from O atom.

\begin{table}[width=.9\linewidth,cols=4,pos=h]
\caption{Coordination numbers of crystalline and glassy Li$_3$ClO.}\label{coord}
\begin{tabular*}{\tblwidth}{@{} CCC@{} }
\toprule
Bond & Crystalline & Amorphous\\
\midrule
O-O & 6 & 4.4\\
O-Li & 6 & 10.25\\
O-Cl & 8 & 4.6\\
Li-Li & 14 & 10.57\\
Li-Cl & 4 & 2.95\\
Cl-Cl & 6 & 3.4\\
\bottomrule
\end{tabular*}
\end{table}

\begin{figure}
    \centering
    \includegraphics[scale=.27]{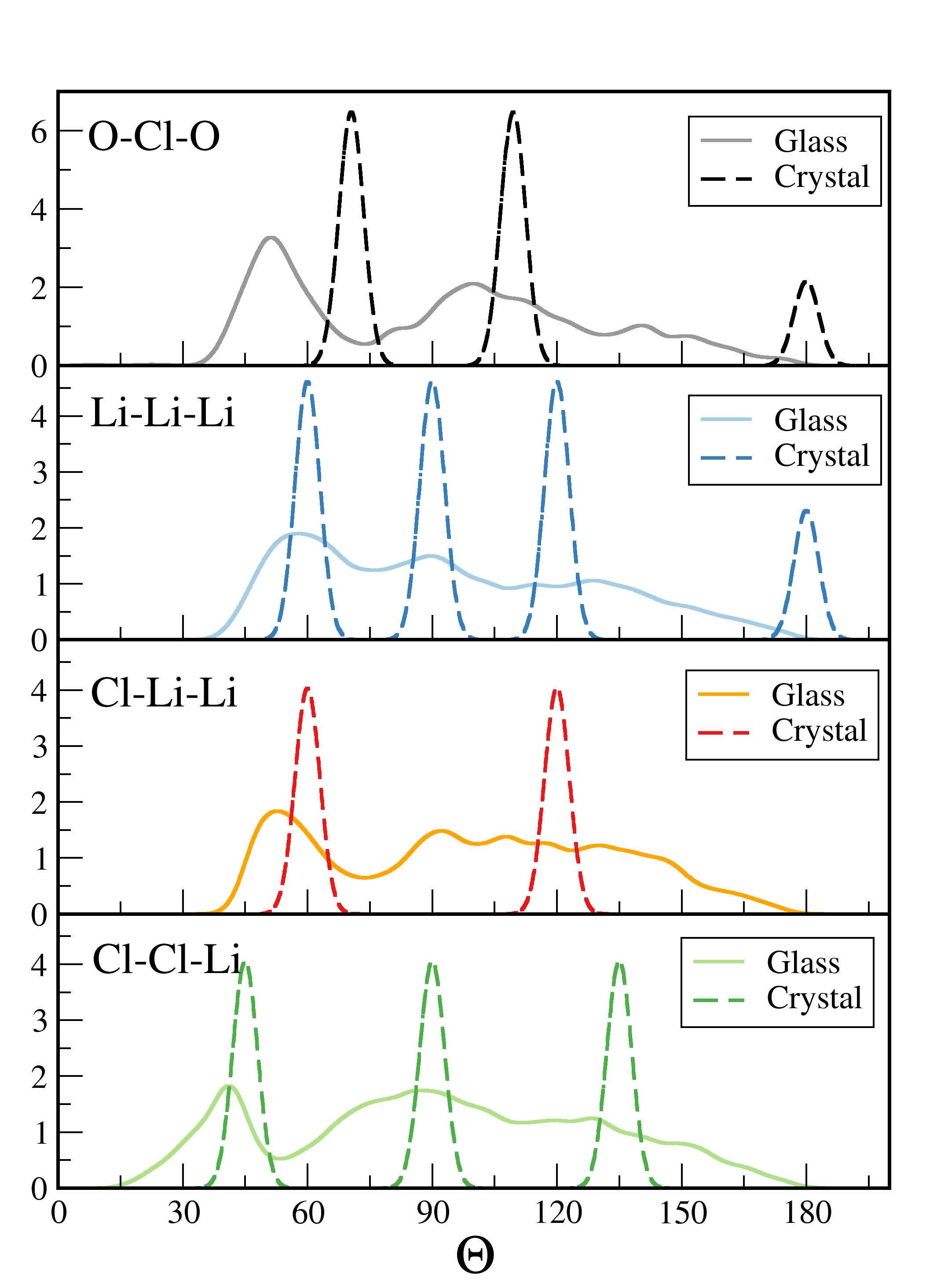}
    \caption{Selected angular distribution function of crystal and amorphous structure of Li$_3$ClO. Full- and dashed-lines
represent amorphous and crystalline structure, respectively. }
    \label{fig:ADF}
\end{figure}

Fig.\ref{fig:ADF} displays the angular distribution function (ADFs) of crystal and amorphous state of Li$_3$ClO. 
It is shown that the O-O-Li angle which is originally located at 0$^{\degree}$ in the crystalline structure 
changes into 32$^{\degree}$ and the O-Li-Li angle which is positioned at 0$^{\degree}$, 45$^{\degree}$, 
and 90$^{\degree}$ changes into 32$^{\degree}$ and 
42$^{\degree}$. Therefore, their angle becomes smaller into an acute angle. On the other hand, 
the O-Cl-Li angle which is located at 36$^{\degree}$ in crystalline state becomes 31$^{\degree}$, 
and thus its change is small compared to other angles presented in the figure.

\subsection{Electronic structure}

\begin{figure}
    \centering
    \includegraphics[scale=.20]{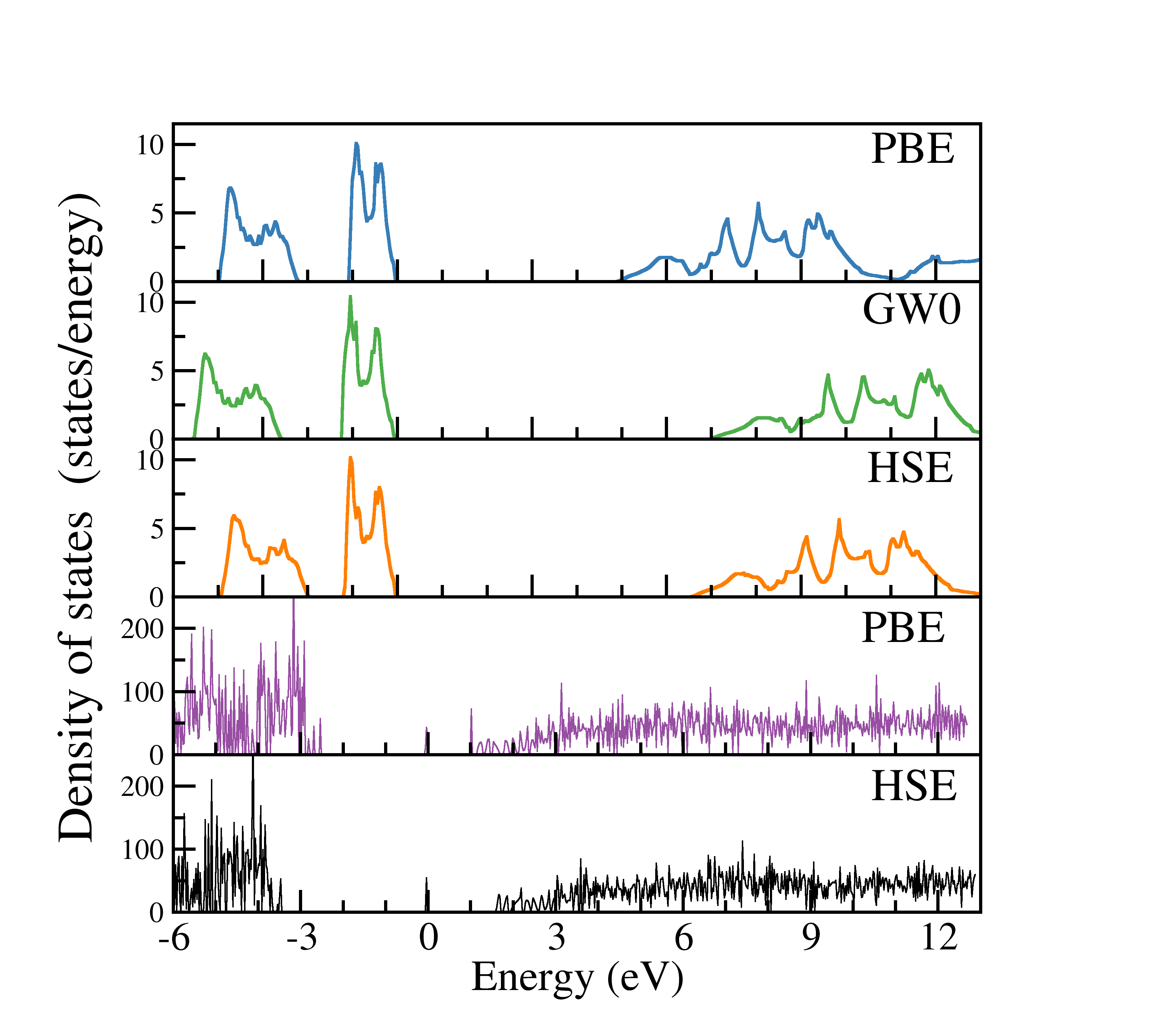}
    \caption{Density of states of crystalline and amorphous Li$_{3}$ClO for several exchange and correlation functionals. 
From top to bottom: crystalline (PBE), crystalline (GW0), crystalline (HSE), glass (PBE), glass (HSE).}
    \label{fig:DOS}
\end{figure}

Fig.~\ref{fig:DOS} shows the density of states (DOS) of crystalline and glassy Li$_{3}$ClO. 
From top to bottom the figure displays DOS for: anti-perovskite Li$_{3}$ClO as calculated by PBE, GW0 and
HSE and glassy Li$_{3}$ClO PBE and HSE. The calculated band gap of crystalline Li$_{3}$ClO as obtained by PBE, 
GW0 and HSE is 4.84 eV, 6.9 eV and 6.48 eV, respectively. These results are in good agreement with other values found in the literature; 6.44~\cite{Braga2014} and 5 eV~\cite{Mo2012}. The band gap of the amorphous system is in comparison relatively smaller,
3.55 eV for PBE and 5.03 for HSE. However we observe that the valence band shifts to higher energy levels.

\section{Conclusions}

We have investigated Li$_{3}$ClO in the amorphous and crystalline state by means of ab initio calculations. Models of the amorphous structure of Li$_{3}$ClO glass were generated by the stochastic quenching method. The amorphous structures were characterized by means of radial distributions functions and angle distribution functions. Coordination numbers and bond lengths were also obtained from the simulations. 
These materials are insulators in both the crystalline state and as a glass. 
The density of states was investigated using PBE, GW0 and the hybrid HSE functionals. As expected the band gap obtained by PBE is smaller than both GW0 and HSE results.

\bibliography{cas-refs}
\bibliographystyle{cas-model2-names}

\end{document}